\documentclass[11pt,a4paper]{article}
\usepackage{jheppub}
\usepackage{bbold} 
\usepackage[pdftex]{pict2e} 
\allowdisplaybreaks[1]

\newcommand{\ba}{\begin{eqnarray}}
\newcommand{\ea}{\end{eqnarray}}
\newcommand{\la}[1]{\label{#1}}
\newcommand{\fig}{figure~}
\newcommand{\eq}{eq.~}
\newcommand{\se}{section~}

\newcommand{\app}{appendix~}
\newcommand{\eqs}{eqs.~}
\newcommand{\nr}[1]{(\ref{#1})}
\newcommand{\ep}{\varepsilon}
\newcommand{\setN}{\mathbb{N}}
\newcommand{\setZ}{\mathbb{Z}}
\newcommand{\sumintL}{L}
\newcommand{\intB}{B}
\newcommand{\Sm}{{\bar H}}
\newcommand{\Sp}{{H}}
\newcommand{\bin}[2]{{#1\choose#2}}
\newcommand{\sumZ}{Z}

\newcommand{\hr}{r}
\newcommand{\po}[2]{\big(#1\big)_{#2}}
\newcommand{\ceil}[1]{\lceil#1\rceil}
\newcommand{\floor}[1]{\lfloor#1\rfloor}

\newcommand{\rmii}[1]{{\mbox{\tiny\rm{#1}}}}
\newcommand{\sumint}[1]{{\hbox{$\sum$}\!\!\!\!\!\!\!\int\,}_{\!\!\!\!\!\raise-0.3ex\hbox{$\scriptstyle{#1}$}}}
\newcommand{\Tinti}[1]{{{\Sigma}\!\!\!\!\raise0.3ex\hbox{$\int$}_\rmii{${#1}$}}}
\newcommand{\mabc}{{m_1,m_2,m_3}}
\newcommand{\nabc}{{\nu_1,\nu_2,\nu_3}}
\newcommand{\eabc}{{\eta_1,\eta_2,\eta_3}}
\newcommand{\Nu}{{\Sigma\nu_i}}
\newcommand{\Eta}{{\Sigma\eta_i}}
\newcommand{\intBasic}{\intB_{1,1,0}^{1,1,0}(d)}
\newcommand{\muf}{{\mu_f}}
\newcommand{\colorA}{magenta}
\newcommand{\colorB}{red}
\newcommand{\colorC}{blue}
\newcommand{\colorD}{black}
\newcommand{\colorE}{cyan}
\renewcommand{\colorA}{black}
\renewcommand{\colorB}{black}
\renewcommand{\colorC}{black}
\renewcommand{\colorD}{black}
\renewcommand{\colorE}{black}

\newcommand{\figA}{%
\begin{figure}
\begin{center}
\begin{picture}(400,100)(0,0)
\put(0,50){\vector(1,0){100}}
\put(50,0){\vector(0,1){100}}
\put(95,41){$n_1$}
\put(35,99){$n_2$}
\multiput(10,10)(10,0){9}{\circle*{4}}
\multiput(10,20)(10,0){9}{\circle*{4}}
\multiput(10,30)(10,0){9}{\circle*{4}}
\multiput(10,40)(10,0){9}{\circle*{4}}
\multiput(10,50)(10,0){9}{\circle*{4}}
\multiput(10,60)(10,0){9}{\circle*{4}}
\multiput(10,70)(10,0){9}{\circle*{4}}
\multiput(10,80)(10,0){9}{\circle*{4}}
\multiput(10,90)(10,0){9}{\circle*{4}}
\put(111,48){$=$}
\put(130,50){\vector(1,0){100}}
\put(180,0){\vector(0,1){100}}
\put(225,41){$n_1$}
\put(165,99){$n_2$}
\multiput(140,50)(10,0){9}{\color{\colorA}\circle*{4}}
\multiput(180,10)(0,10){9}{\color{\colorC}\circle*{4}}
\multiput(140,10)(10,10){9}{\color{\colorB}\circle*{4}}
\multiput(140,90)(10,-10){9}{\circle*{4}}
\put(180,50){\color{\colorE}\circle*{4}}
\put(180,50){\circle{4}}
\put(241,48){$+$}
\put(260,50){\vector(1,0){100}}
\put(310,0){\vector(0,1){100}}
\put(355,41){$n_1$}
\put(295,99){$n_2$}
\color{\colorA}
\put(270,20){\circle*{4}}
\put(270,30){\circle*{4}}
\put(270,40){\circle*{4}}
\put(280,30){\circle*{4}}
\put(280,40){\circle*{4}}
\put(290,40){\circle*{4}}
\put(330,60){\circle*{4}}
\put(340,60){\circle*{4}}
\put(340,70){\circle*{4}}
\put(350,60){\circle*{4}}
\put(350,70){\circle*{4}}
\put(350,80){\circle*{4}}
\color{\colorB}
\put(280,10){\circle*{4}}
\put(290,10){\circle*{4}}
\put(290,20){\circle*{4}}
\put(300,10){\circle*{4}}
\put(300,20){\circle*{4}}
\put(300,30){\circle*{4}}
\put(320,70){\circle*{4}}
\put(320,80){\circle*{4}}
\put(320,90){\circle*{4}}
\put(330,80){\circle*{4}}
\put(330,90){\circle*{4}}
\put(340,90){\circle*{4}}
\color{\colorC}
\put(280,90){\circle*{4}}
\put(290,90){\circle*{4}}
\put(290,80){\circle*{4}}
\put(300,90){\circle*{4}}
\put(300,80){\circle*{4}}
\put(300,70){\circle*{4}}
\put(320,10){\circle*{4}}
\put(320,20){\circle*{4}}
\put(320,30){\circle*{4}}
\put(330,10){\circle*{4}}
\put(330,20){\circle*{4}}
\put(340,10){\circle*{4}}
\color{\colorD}
\put(270,60){\circle*{4}}
\put(270,70){\circle*{4}}
\put(270,80){\circle*{4}}
\put(280,60){\circle*{4}}
\put(280,70){\circle*{4}}
\put(290,60){\circle*{4}}
\put(330,40){\circle*{4}}
\put(340,40){\circle*{4}}
\put(340,30){\circle*{4}}
\put(350,40){\circle*{4}}
\put(350,30){\circle*{4}}
\put(350,20){\circle*{4}}
\end{picture}
\caption{\label{figA} Different regions in the Matsubara double-sum, as used for deriving \eq\nr{eq:pos}. 
The regions in the first diagram on the right are one-dimensional sums only and can be immediately written as 
Riemann zeta functions, while we choose to map all eight regions of the last diagram to the 
{\color{\colorA} lower-right region in the first quadrant}, i.e.\ to double-sums of type $\sum_{n_1>n_2>0}$.}
\end{center}
\end{figure}}

\newcommand{\figB}{%
\begin{figure}
\begin{center}
\begin{picture}(400,70)(0,0)
\put(10,10){\vector(1,0){50}}
\put(10,10){\vector(0,1){50}}
\put(55,1){$n_1$}
\put(-5,59){$n_2$}
\put(20,20){\circle*{4}}
\put(30,20){\circle*{4}}
\put(40,20){\circle*{4}}
\put(50,20){\circle*{4}}
\put(20,30){\circle*{4}}
\put(30,30){\circle*{4}}
\put(40,30){\circle*{4}}
\put(50,30){\circle*{4}}
\put(20,40){\circle*{4}}
\put(30,40){\circle*{4}}
\put(40,40){\circle*{4}}
\put(50,40){\circle*{4}}
\put(20,50){\circle*{4}}
\put(30,50){\circle*{4}}
\put(40,50){\circle*{4}}
\put(50,50){\circle*{4}}
\put(76,28){$=$}
\put(110,10){\vector(1,0){50}}
\put(110,10){\vector(0,1){50}}
\put(155,1){$n_1$}
\put(95,59){$n_2$}
\put(130,20){\circle*{4}}
\put(140,20){\circle*{4}}
\put(150,20){\circle*{4}}
\put(140,30){\circle*{4}}
\put(150,30){\circle*{4}}
\put(150,40){\circle*{4}}
\put(176,28){$+$}
\put(210,10){\vector(1,0){50}}
\put(210,10){\vector(0,1){50}}
\put(255,1){$n_1$}
\put(195,59){$n_2$}
\put(220,30){\circle*{4}}
\put(220,40){\circle*{4}}
\put(230,40){\circle*{4}}
\put(220,50){\circle*{4}}
\put(230,50){\circle*{4}}
\put(240,50){\circle*{4}}
\put(276,28){$+$}
\put(310,10){\vector(1,0){50}}
\put(310,10){\vector(0,1){50}}
\put(355,1){$n_1$}
\put(295,59){$n_2$}
\put(320,20){\circle*{4}}
\put(330,30){\circle*{4}}
\put(340,40){\circle*{4}}
\put(350,50){\circle*{4}}
\end{picture}
\caption{\label{figB} For the proof of the shuffle relation \eq\nr{eq:shuffle}, the sum over all pairs of positive integers in the first quadrant 
is replaced by the sum over the three regions on the right-hand side.}
\end{center}
\end{figure}}

\title{Factorizing two-loop vacuum sum-integrals}

\author[a,b]{Andrei~I.~Davydychev,}
\author[a,c]{Pablo~Navarrete}
\author[a]{and York~Schr\"oder}

\affiliation[a]{Centro de Ciencias Exactas, Departamento de Ciencias B\'asicas, Universidad del B\'io-B\'io, Avenida Andr\'es Bello 720, Chill\'an, Chile}
\affiliation[b]{Grenfell Campus, Memorial University of Newfoundland, 20 University Dr., Corner Brook, NL, A2H 5G4, Canada}
\affiliation[c]{Department of Physics and Helsinki Institute of Physics, P.O.\ Box 64, University of Helsinki, Finland}

\emailAdd{adavyd@ubiobio.cl}
\emailAdd{pablo.navarrete@helsinki.fi}
\emailAdd{yschroder@ubiobio.cl}

\arxivnumber{2312.17367}

\abstract{We derive analytic results for scalar massless bosonic vacuum sum-integrals at two loops. 
Building upon a recent factorization proof of massive two-loop vacuum integrals, we are able to solve the 
corresponding Matsubara sums and map the result onto one-loop structures, 
thereby proving factorization also in the sum-integral setting.
Analytic results are provided for generic integer-valued propagator- and numerator-powers of the class of
sum-integrals under consideration, allowing to eliminate them from any perturbative expansion, dramatically
simplifying the evaluation of some observables encountered e.g.\ in hot QCD.}

\begin{document}
\maketitle

%
\section{Introduction}
\la{se:intro}

Due to their phenomenological importance in early-universe cosmology, compact-star astrophysics, and heavy-ion collisions,
substantial effort has been spent on evaluations of quantities relevant
in hot and/or dense systems in and near equilibrium. 
The main emphasis has been on understanding quantum chromodynamics (QCD) \cite{Gross:2022hyw} 
or generalizations thereof in such extreme conditions,
since the interplay of asymptotic freedom at high energies (temperatures $T$ and/or densities as defined by 
the fermion chemical potentials $\muf$) and strong coupling at low energies 
provides for a very rich scenario, in which an interplay of different theoretical methods becomes important.

Within finite-temperature quantum field theory (see \cite{Kapusta:2023eix,Bellac:2011kqa,Laine:2016hma,Ghiglieri:2020dpq}
for textbook-level treatments), the evaluation of vacuum-type integrals plays a central role 
in the determination of equilibrium observables, such as the equation of state (EoS) of a thermal system \cite{Arnold:1994ps,Arnold:1994eb}.
The EoS is typically given in terms of the pressure (or free energy) of the system under consideration, 
and constitutes a crucial ingredient in data analysis and modeling in the phenomenological applications mentioned above.
First-principles determinations of the QCD EoS via Monte Carlo simulations of lattice gauge theory 
\cite{Boyd:1996bx,Borsanyi:2012ve,Giusti:2016iqr} are 
difficult in particular when fermions are present, leaving weak-coupling expansions as a viable alternative.

One very fruitful weak-coupling approach, for example, has been the application of effective field theory (EFT) 
methods to hot and dense systems,
in particular in the non-abelian setting of QCD, where the gauge symmetry severely constrains (but also guides)
methods that go beyond naive perturbative expansions\footnote{Another approach to account for low-energy screening effects 
in a gauge theory is {\em hard-thermal-loop perturbation theory}
\cite{Braaten:1989kk,Braaten:1989mz,Andersen:1999fw}, 
which has been developed up to and including three-loop contributions \cite{Haque:2014rua,Andersen:2015eoa},
superseding simpler resummation schemes that had been developed for scalar theories, 
such as {\em screened perturbation theory} \cite{Karsch:1997gj}.}. 
This has led to the development and application of dimensionally reduced EFT's 
\cite{Ginsparg:1980ef,Appelquist:1981vg,Kajantie:1995dw,Braaten:1995jr} that describe the strongly coupled
low-energy QCD degrees of freedom in a systematic way, taking account of the correct high-energy behavior 
by mapping all EFT couplings onto QCD parameters (strong gauge coupling, temperature, 
group structure and field content, fermion masses and chemical potentials).
Sometimes, and in particular for bosonic systems, comparison with lattice simulations are possible and provide 
valuable insight into questions such as convergence and range of validity of perturbative series. 

For the QCD pressure, for example, almost all contributions up to the order
of $g^6$ (in a formal power-counting in terms of the gauge coupling $g$, 
corresponding to the four-loop level in a diagrammatic expansion) that arise
from QCD as well as two levels of EFT's that correspond to screened and unscreened `soft' gluonic modes
are known \cite{Kajantie:2002wa,Vuorinen:2003fs,Gorda:2023mkk}, 
except for a well-defined but technically challenging perturbative piece from the `hard' QCD sector.
While 3d lattice simulations already play a role in determining parts of the low-energy contributions \cite{DiRenzo:2006nh},
it has even been possible to obtain an order-of-magnitude estimate for the missing perturbative piece 
from comparisons with 4d lattice simulations at intermediate temperatures where both, weak-coupling and lattice approaches
are justified \cite{Laine:2006cp,Schroder:2006vz}.
From a theoretical standpoint this is not a satisfactory situation, however, strongly motivating the development of 
techniques that allow for an evaluation of the missing perturbative contributions, see \cite{Navarrete:2022adz} for
the state-of-the-art at $\muf=0$, and \cite{Gorda:2023mkk} for $\muf\neq 0$.

The aim of the present paper is to advance perturbative techniques that directly apply to such contributions,
necessitating the evaluation of many sum-integrals (for a definition, see \se\ref{se:notation} below).
As has been repeatedly observed in the past, many two-loop sum-integrals can be decomposed into one-loop factors
that are known in terms of Zeta and Gamma functions, 
allowing for analytic solutions in the space-time dimension $D=d+1$. See \se\ref{se:checks} for a number of known examples
from the literature.
In the present paper, we prove that this decomposition is generic, and give an algorithm that constructs this 
decomposition for any scalar massless bosonic two-loop vacuum sum-integral.

In the fermionic case, the situation is less clear. While again some isolated two-loop sum-integral factorization formulas 
have been observed in the literature, when both scales $T$ and $\muf$ are present a recursive treatment might still
apply to reduce the generic two-loop case to one-loop factors \cite{Osterman:2023tnt}. 

The structure of the paper is as follows. 
After introducing some basic notation in \se\ref{se:notation}, 
we introduce a generic massless two-loop vacuum sum-integral as the main object of our paper in \se\ref{se:setup},
and immediately represent it as infinite double (Matsubara-) sums over of massive two-loop (continuum-) integrals.
For the latter, for generic positive values of propagator powers and in the specific mass-combinations of interest here, 
a factorization formula is known, whose details we recall in \se\ref{se:A}.
In \se\ref{se:C}, we then show how to manipulate the double sums such as to generalize the factorization formula 
to the sum-integral setting, via four separate lines of proof. 
We finally present our factorized result as well as an extension to non-positive propagator powers in \se\ref{se:result},
and close with a brief discussion of possible generalizations in \se\ref{se:conclu}. 
Two short appendices collect details about single and double sums of interest, 
as well as sample Mathematica code that implements our reduction algorithm.

%
\section{Notation and one-loop example}
\la{se:notation}

Let us start with some definitions, to set the stage for the rather technical nature of the present paper.
Our setting is thermal field theory in the imaginary-time formalism, defined on a $(d+1)$ dimensional Euclidean space 
with a compact temporal coordinate $\tau$ (which emerges after Wick rotating $it\rightarrow\tau$, 
leading to $e^{iS}\rightarrow e^{-S_{\rm E}}$ in the path integral) 
that has period $1/T$ set by the inverse temperature. 
Dimensionally regularized integrals then employ $d=3-2\ep$ in our conventions. 
Since we do not $\ep$-expand but work fully analytically in this paper, the particular integer value of the space-time dimension will not play any role below.

Bosonic (fermionic) fields are periodic (antiperiodic) functions of $\tau$.
This leads to discrete temporal Fourier transforms to momentum space, such that momentum integrals are traded for sum-integrals.
We work with Euclidean four-momenta $P=(P_0,p)$ with bosonic\footnote{For fermions one would have $P_0 = (2n_p + 1)\pi T+i\muf$ 
where $\muf$ is the chemical potential of the respective fermion flavor.
In this work we focus on the bosonic case, however, and only briefly comment on fermionic generalizations in the conclusions.}
Matsubara frequencies  $P_0=2n_p\pi T$ where $n_p\in\setZ$.
Massless propagators are $1/[P^2]=1/[(P_0)^2+p^2]$, and we employ the sum-integral symbol defined as
\ba
\sumint{P} \equiv T\sum_{n_p\in\setZ} \int\frac{{\rm d}^dp}{(2\pi)^d} \;.
\ea

While some of the machinery used in organizing and reducing perturbative expansions, such as e.g.\ integration-by-parts (IBP) 
methods \cite{Chetyrkin:1981qh,Laporta:2000dsw},
carries over from zero to finite temperatures \cite{Nishimura:2012ee}, 
evaluating master integrals is severely complicated by the Matsubara sums encountered
at finite $T$. In particular, the number of analytically known sum-integrals (mostly simple one- and two-loop cases) is rather limited, 
while a number of phenomenologically relevant cases at three and four loops have been worked out up to a couple of terms in 
their $\ep$ expansion, typically relying on tedious subtractions of sub-divergences in one-loop substructures \cite{Arnold:1994ps}, 
and resorting to numerical evaluations for the constant terms.

As one concrete well-known example for an analytic solution, let us warm up by discussing the somewhat trivial 
1-loop massless bosonic vacuum sum-integrals ($\eta\in\setN_0$)
\ba \la{eq:1loopDef}
I_\nu^\eta(d,T) &\equiv& \sumint{P} \frac{(P_0)^\eta}{[P^2]^\nu} \;.
\ea
One way to evaluate them is to view the $(d+1)$-dimensional sum-integral with massless propagators 
as a sum over massive $d$-dimensional 1-loop tadpole integrals ($m\ge0$)
\ba \la{eq:1loopJ}
J_\nu(d,m) &\equiv& \int\frac{{\rm d}^dp}{(2\pi)^d}\frac1{[m^2+p^2]^\nu} 
\;=\; \big[m^2\big]^{\frac{d}2-\nu}\frac{\Gamma(\nu-\tfrac{d}2)}{(4\pi)^{d/2}\,\Gamma(\nu)} \;.
\ea
Taking advantage of the simple mass dependence of the continuum integral $J$, this gives
\ba
I_\nu^\eta(d,T) &=& T\sum_{n\in\setZ} (2n\pi T)^\eta \,J_\nu(d,|2n\pi T|) 
\nonumber\\&=&  T\delta_{\eta,0} J_\nu(0) + \big[1+(-1)^\eta\big]T\sum_{n=1}^\infty (2n\pi T)^\eta J_\nu(d,2n\pi T)
\nonumber\\&=& 0+\big[1+(-1)^\eta\big]T\sum_{n=1}^\infty (2n\pi T)^{\eta+d-2\nu} \, J_\nu(d,1)
\nonumber\\&=&  \frac{[1+(-1)^\eta]\,T\,\zeta(2\nu-\eta-d)}{(2\pi T)^{2\nu-\eta-d}}\,\frac{\Gamma(\nu-\tfrac{d}2)}{(4\pi)^{d/2}\,\Gamma(\nu)} \;, \la{eq:1loop}
\ea
where we have first separated the point $n=0$ where propagator mass vanishes 
(leaving a massless tadpole that vanishes identically in dimensional regularization),
mapped negative Matsubara indices onto positive ones, 
and used \eq\nr{eq:1loopJ} to scale out the mass to finally identify the single Matsubara sum as a zeta function.
As it should be, due to the sum-integral's symmetry under a momentum shift $P\rightarrow -P$, all $I_\nu^{\eta_{\rm odd}}(d,T)$ vanish identically,
as is reflected by the prefactor $[1+(-1)^\eta]$. Therefore, we only encounter zeta values of the form $\zeta(n_{\rm even}-d)$
at one loop\footnote{At finite $\muf$, the symmetry argument breaks down, 
such that the corresponding one-loop tadpoles do not vanish for odd values of $\eta$. Furthermore, at finite $T$ and $\muf$ one encounters 
Hurwitz (instead of Riemann) zeta functions.}, 
a fact that will become important to recall in \se\ref{se:result}.

The goal of the present paper is to expand the set of known generic-index analytic solutions to the two-loop level,
going beyond the finite number of fixed-index cases that had been solved by IBP methods in the past.
In analogy to the 1-loop case above, the main idea is to view a massless multi-loop sum-integral as a multiple sum over 
massive $d$-dimensional momentum integrals, where Matsubara frequencies such as $P_0$ play the role of propagator masses.

%
\section{Massless bosonic two-loop vacuum sum-integrals: Setup}
\la{se:setup}

The general scalar massless bosonic 2-loop vacuum sum-integral is defined as 
\ba\la{eq:L}
\sumintL_\nabc^\eabc(d,T) &\equiv& 
\sumint{PQ} \frac{(P_0)^{\eta_1}\,(Q_0)^{\eta_2}\,(P_0-Q_0)^{\eta_3}}{[P^2]^{\nu_1}\,[Q^2]^{\nu_2}\,[(P-Q)^2]^{\nu_3}}
\;.
\ea
Imitating the strategy presented in the evaluation of \eq\nr{eq:1loop} above, 
one can view the massless two-loop sum-integral $\sumintL$ as a double sum over massive continuum integrals $\intB$, 
with the Matsubara frequencies playing the role of the continuum integral's masses, as
\ba \la{eq:LB}
\sumintL_\nabc^\eabc(d,T) &=& 
\frac{T^2}{(2\pi T)^{2\Nu-\Eta-2d}}
\sum_{n_1,n_2\in \setZ}
n_1^{\eta_1}\,n_2^{\eta_2}\,(n_1-n_2)^{\eta_3}\,
\intB^\nabc_{n_1,n_2,n_1-n_2}(d) \;,\\
\la{eq:defB}
\intB^\nabc_\mabc(d) &\equiv&
\int\!\!\frac{{\rm d}^dp}{(2\pi)^d}
\int\!\!\frac{{\rm d}^dq}{(2\pi)^d}
\frac{1}{[m_1^2+p^2]^{\nu_1}\,[m_2^2+q^2]^{\nu_2}\,[m_3^2+(p-q)^2]^{\nu_3}} \;,
\ea
where $\Nu\equiv\nu_1+\nu_2+\nu_3$ and $\Eta\equiv\eta_1+\eta_2+\eta_3$.
In general, one would expect the massive integrals $\intB$ to have a complicated dependence 
on (ratios of) its three masses. In fact, general results in terms of Appell’s hypergeometric function $F_4$ 
can be found in \cite{Davydychev:1992mt}.
In our particular case, however, these masses are seen to be linearly related, 
due to 4-momentum conservation at the sum-integral's vertices. 
For linearly related masses there are significant simplifications, as recently pointed out in \cite{Davydychev:2022dcw} 
by a detailed analysis of the corresponding IBP relations, leading to a general factorization formula of the type $\intB\rightarrow \sum J\cdot J$.
The idea is now to use results of \cite{Davydychev:2022dcw} for the massive 3d vacuum tadpole $\intB$ of \eq\nr{eq:LB}, 
and do the double-sum afterwards. 
Amazingly, this will allow us to present a proof of factorization for, schematically, $\sumintL\rightarrow \sum I\cdot I$, 
where the $I$ are the 1-loop sum-integrals of \eq\nr{eq:1loop}. 

Further following the successful 1-loop strategy as explained below \eq\nr{eq:1loop}, 
let us start by sorting out the cases where masses of $\intB$ can vanish. 
To this end, we decompose the double-sum into sectors where the masses are always positive 
(see \fig\ref{figA} for region mapping), taking into account that $\intB$ depends on squared masses only, 
and use the integral's symmetry to re-order some mass-index pairs in $\intB$. We obtain\footnote{Using the shortcut 
$\delta_j\equiv \delta_{j,0}$ for the Kronecker delta function.}
\ba \la{eq:pos}
\sumintL_\nabc^\eabc(d,T) &=& 
\frac{T^2\,[1+(-1)^\Eta]}{(2\pi T)^{2\Nu-\Eta-2d}}\, \Big\{
\tfrac12\,\delta_{\eta_1} \delta_{\eta_2} \delta_{\eta_3} \,{\color{\colorE}\intB_{0,0,0}^\nabc}
+\zeta(2\Nu-\Eta-2d)\!\times
\nonumber\\&&{}
\times\!\Big[
(-1)^{\eta_3}\delta_{\eta_1}{\color{\colorC}\intB_{0,1,1}^\nabc}
+\delta_{\eta_2}{\color{\colorA}\intB_{0,1,1}^{\nu_2,\nu_1,\nu_3}}
+\delta_{\eta_3}{\color{\colorB}\intB_{0,1,1}^{\nu_3,\nu_1,\nu_2}}
+2^{\eta_3}(-1)^{\eta_2}{\color{\colorD}\intB_{1,1,2}^\nabc}
\Big]
\nonumber\\&&{}
+{\color{\colorA}\Sm_{\eta_3,\eta_2,\eta_1}^{\nu_3,\nu_2,\nu_1}}
+(-1)^{\eta_3} {\color{\colorB}\Sm_{\eta_3,\eta_1,\eta_2}^{\nu_3,\nu_1,\nu_2}}
+(-1)^{\eta_2} {\color{\colorC}\Sp_{\eta_2,\eta_1,\eta_3}^{\nu_2,\nu_1,\nu_3}}
+(-1)^{\eta_2} {\color{\colorD}\Sp_\eabc^\nabc}
\Big\} \;,\qquad\\
\la{eq:Sp}
\mbox{with}&&\Sp_{\eta_a,\eta_b,\eta_c}^{\nu_a,\nu_b,\nu_c}(d) \equiv \sum_{n_1>n_2>0} 
n_1^{\eta_a} n_2^{\eta_b} (n_1+n_2)^{\eta_c} \, \intB_{n_1,n_2,n_1+n_2}^{\nu_a,\nu_b,\nu_c}(d) \;,\\
\la{eq:Sm}
\mbox{and}&&\Sm_{\eta_a,\eta_b,\eta_c}^{\nu_a,\nu_b,\nu_c}(d) \equiv \sum_{n_1>n_2>0} 
(n_1-n_2)^{\eta_a} n_2^{\eta_b} n_1^{\eta_c} \, \intB_{n_1-n_2,n_2,n_1}^{\nu_a,\nu_b,\nu_c}(d) \;.
\ea
Naturally, the sum-integral $\sumintL$ vanishes for odd $\Eta$, as is immediately clear from the definition \eq\nr{eq:L}, 
considering the integrand's symmetry under the simultaneous shifts $P\rightarrow-P$ and $Q\rightarrow-Q$.
The special-mass cases $\intB^\nabc_{0,0,0}(d)$ and $\intB^\nabc_{0,1,1}(d)$ are known analytically in $d$ 
dimensions \cite{Vladimirov:1979zm},
\ba
\intB_{0,0,0}^\nabc(d) &=& 0 \mbox{~(massless tadpoles vanish in dimensional regularization)}\;,\\
\intB_{0,1,1}^\nabc(d) &=& \frac{\Gamma(\nu_1+\nu_2-\frac{d}2)\Gamma(\nu_1+\nu_3-\frac{d}2)\Gamma(\frac{d}2-\nu_1)\Gamma(\Nu-d)}
{\Gamma(\nu_2)\Gamma(\nu_3)\Gamma^2(1-\frac{d}2)\Gamma(\frac{d}2)\Gamma(\nu_1+\Nu-d)}\,\intBasic  \\
\mbox{with} && \intBasic \equiv \frac{\Gamma^2(1-\frac{d}2)}{(4\pi)^d} = \big[J_1(d,1)\big]^2 
\quad \mbox{(see \eq\nr{eq:1loopJ})} \;, \la{eq:BnormJ}
\ea
while $\intB_{1,1,2}^\nabc(d)$ can be extracted from \cite{Davydychev:1992mt} (cf.\ \app C of \cite{Davydychev:2022dcw}). 
Note however that these specific results are not necessarily relevant for our treatment of $\sumintL$, as can be appreciated 
from our final result \eq\nr{eq:final} below.
\figA

%
\section{Massive two-loop vacuum integrals: Factorization formula}
\la{se:A}

Let us now examine the massive two-loop scalar vacuum integrals $\intB^\nabc_\mabc(d)$ that were defined in \eq\nr{eq:defB}, 
over whose masses we need to sum. 
As already mentioned above, general results in terms of Appell's hypergeometric function $F_4$ of two variables (mass ratios) 
can be found in \eq(4.3) of \cite{Davydychev:1992mt}.
This is not very practical for our purpose, since it involves four infinite sums. 
One can actually do (much) better by analyzing the integrals $\intB$ from scratch, at the special kinematic point 
that is of relevance here. Stemming from the Matsubara sums, in the case of interest here the masses obey a `sum-rule' at each vertex; 
in fact, according to the mapping in \eq\nr{eq:pos} we only need $m_3=m_1+m_2$. For this special ``collinear'' case, 
the general results of \cite{Davydychev:1992mt} can be simplified considerably.

Imposing the linear relation between the three masses, one can set up an integration-by-parts (IBP) recursion that reduces the integral $\intB$ 
for all positive values of $\nu_i\in \setN$ 
towards one or zero. Relevant explicit formulas can be found e.g.\ in \eqs(92,93,97) of \cite{Tarasov:1997kx}. 
We have recently been able to solve these IBP recursion relations 
explicitly \cite{Davydychev:2022dcw}, and obtained (from now on we implicitly assume $m_3=m_1+m_2$ in all equations; 
also, $\Nu=\nu_1+\nu_2+\nu_3$)
\ba \la{eq:newconj}
\intB^\nabc_\mabc(d) &=& 
\intBasic\, \Bigg\{
\sum_{j=1-\nu_1}^{\nu_2-1} (-1)^\Nu \, c^{(\Nu)}_{\nu_1,\nu_2;j}(d) \, m_1^{d-\Nu+j} \, m_2^{d-\Nu-j}
\nonumber\\&&~~~~~+
\sum_{j=1-\nu_1}^{\nu_3-1} (-1)^j \, c^{(\Nu)}_{\nu_1,\nu_3;j}(d) \, m_1^{d-\Nu+j} \, m_3^{d-\Nu-j}
\nonumber\\&&~~~~~+
\sum_{j=1-\nu_2}^{\nu_3-1} (-1)^j \, c^{(\Nu)}_{\nu_2,\nu_3;j}(d) \, m_2^{d-\Nu+j} \, m_3^{d-\Nu-j} 
\;\Bigg\} \Bigg|_{\scriptsize m_3=m_1+m_2} \;,
\ea
where the coefficients $c^{(\Nu)}_{\nu_a,\nu_b;j}(d)$ are rational functions in $d$ that obey a number 
of symmetries, such as
\ba \la{eq:csy}
c^{(\Nu)}_{\nu_a,\nu_b;j}(d) &=& c^{(\Nu)}_{\nu_b,\nu_a;-j}(d) \;,\quad
c^{(\Nu)}_{\nu_a,\nu_b;0}(d) = c^{(\Nu)}_{\nu_b,\nu_a;0}(d) \;,
\ea
and are given explicitly by \cite{Davydychev:2022dcw} 
\ba\la{eq:kallen3}
c^{(\Nu)}_{\nu_a,\nu_b;j}(d) &=& 
\frac{(-1)^{\Nu-n_j+1}\,\po{1-\frac{d}2}{n_j-j-1}}{2\,\po{\frac12}{n_j-\nu_b-\nu_c}\,\po{\frac12}{n_j-j-\nu_a-\nu_c}\,(\nu_c-1)!}
\times\\&\times& 
\!\!\!\!\!\!\sum_{k={\rm max}(1\!+\!j,1)}^{{\rm min}(\nu_a\!+\!j,\nu_b)} 
\frac{\po{\frac{d}2-n_j+1}{k-1}\,(n_j-k-1)!}{\po{\frac{d+3}2\!-\!\Nu}{n_j-k} \po{\frac12}{\nu_c-n_j+k}\,(k\!-\!1)!\,(k\!-\!j\!-\!1)!\,(\nu_b\!-\!k)!\,(\nu_a\!-\!k\!+\!j)!}
\nonumber
\ea
with Pochhammer symbols $(a)_\nu\equiv\frac{\Gamma(a+\nu)}{\Gamma(a)}$ and integers $n_j=\ceil{\frac{\Nu+j}2}$. 
Note that using \eq\nr{eq:newconj} in \eq\nr{eq:pos} explicitly decomposes the two-loop sum-integral $\sumintL$ 
into a sum over one-loop factors, a fact that we will exploit in \se\ref{se:C} below.

For later use, we spell out \eq\nr{eq:newconj} for two special cases that appear in \eq\nr{eq:pos}: 
\ba \la{eq:new0mm}
\intB_{0,m,m}^\nabc(d) &=& m^{2d-2\Nu} \, \intBasic\,
\Big\{\beta^\nabc(d)\Big\}\;,\\
\mbox{with}&&\beta^\nabc(d) =
\sum_{j=1-\nu_2}^{\nu_3-1} (-1)^j\,c^{(\Nu)}_{\nu_2,\nu_3;j}(d) \;,\\
\la{eq:newmm2m}
\intB_{m,m,2m}^\nabc(d) &=& m^{2d-2\Nu} \, \intBasic\,
\Big\{ \hr_1^\nabc(d) + 2^{d-2}\,\hr_2^\nabc(d) \Big\} \;,\\
\la{eq:newr32}
\mbox{with}&&\hr_1^\nabc(d)=(-1)^\Nu \sum_{j=1-\nu_1}^{\nu_2-1} c^{(\Nu)}_{\nu_1,\nu_2;j}(d) \;,\\
\mbox{and}&&\hr_2^\nabc(d)=2^{2-\Nu}\Big[
\sum_{j=1-\nu_1}^{\nu_3-1} (-\tfrac12)^j\,c^{(\Nu)}_{\nu_1,\nu_3;j}(d)
+\sum_{j=1-\nu_2}^{\nu_3-1} (-\tfrac12)^j\,c^{(\Nu)}_{\nu_2,\nu_3;j}(d) \Big] \,.\qquad
\ea

%
\section{Massless bosonic two-loop vacuum sum-integrals: Matsubara sums}
\la{se:C}

The final task is now to evaluate the remaining double sums $\Sp$ and $\Sm$ in \eq\nr{eq:pos}. 
Having \eq\nr{eq:newconj} at hand, the mass structure is fixed explicitly, and we can work on performing the sums without 
actually specifying the rational coefficient functions $c^{(\Nu)}_{\nu_a,\nu_b;j}(d)$ yet. 

We will now show how the specific linear combination of sums $\Sp$ and $\Sm$ that occurs in \eq\nr{eq:pos} combine to 
\begin{itemize}
\item[(a)] cancel all $\sumZ$ (for a definition, see \eq\nr{eq:doublesum4}) in the last two terms of \eq\nr{eq:pos}, 
\item[(b)] cancel all $\zeta(i,j)$ (see \eq\nr{eq:doublesum1}) in the sum of all four terms of the last line of \eq\nr{eq:pos}, 
\item[(c)] cancel all remaining single $\zeta(i)$ in \eq\nr{eq:pos} to 
\item[(d)] leave us with the expected tadpole reductions that contain products $\zeta(i)\,\zeta(j)$ containing only $\zeta(n_{\rm even}-d)$
allowing us to map onto the 1-loop sum-integrals of \eq\nr{eq:1loop}, 
\end{itemize}
all of which allows us to formulate our final result in a compact way.

%
\subsection{Proof of statement (a)}
\la{se:proofA}

Plugging \eq\nr{eq:newconj} into \eq\nr{eq:Sp}, re-expressing the `unwanted mass' (the one not already appearing in the $j$-sum)
in the prefactor in terms of the two others 
(by introducing an additional binomial sum, as for example in 
$n_1^{\eta_1}=\sum_{k=0}^{\eta_1}\bin{\eta_1}{k}(-n_2)^k(n_1+n_2)^{\eta_1-k}$), 
shifting summation indices appropriately (e.g.\ $k\rightarrow\eta_3-k$ and $j\rightarrow-j$) 
and using the symmetry \eq\nr{eq:csy} to collect similar expressions, the last two terms of \eq\nr{eq:pos} combine to
\ba \la{eq:HH}
&&\frac{\Sp_{\eta_2\eta_1\eta_3}^{\nu_2\nu_1\nu_3}+\Sp_{\eta_1\eta_2\eta_3}^{\nu_1\nu_2\nu_3}}{\intBasic}
\;=\\&&\hphantom{+}
\sum_{j=1-\nu_1}^{\nu_2-1} (-1)^\Nu\,c^{(\Nu)}_{\nu_1,\nu_2;j}(d) 
\sum_{k=0}^{\eta_3} \bin{\eta_3}{k} 
\sum_{n_1>n_2>0}\Big(n_2^{d-\ell_1}n_1^{d-2\Nu+\Eta+\ell_1}+n_1^{d-\ell_1}n_2^{d-2\Nu+\Eta+\ell_1}\Big)
\nonumber\\&&+
\sum_{j=1-\nu_1}^{\nu_3-1} (-1)^j\,c^{(\Nu)}_{\nu_1,\nu_3;j}(d) 
\sum_{k=0}^{\eta_2} \bin{\eta_2}{k} (-1)^{k}
\sum_{n_1>n_2>0}\Big(n_1^{d-\ell_1}+n_2^{d-\ell_1}\Big)
\Big(n_1+n_2\Big)^{d-2\Nu+\Eta+\ell_1}
\nonumber\\&&+
\sum_{j=1-\nu_2}^{\nu_3-1} (-1)^j\,c^{(\Nu)}_{\nu_2,\nu_3;j}(d) 
\sum_{k=0}^{\eta_1} \bin{\eta_1}{k} (-1)^{k}
\sum_{n_1>n_2>0}\Big(n_1^{d-\ell_2}+n_2^{d-\ell_2}\Big)
\Big(n_1+n_2\Big)^{d-2\Nu+\Eta+\ell_2} \nonumber
\ea
where we have abbreviated $\ell_p\equiv \Nu-\eta_p-j-k$, and the $\bin{\eta_p}{k}$ are binomial coefficients.
The double sums in the last two lines of \eq\nr{eq:HH} have the form
\ba
&&\sum_{n_1>n_2>0}\Big(\frac1{n_1^\alpha}+\frac1{n_2^\alpha}\Big)\frac1{(n_1+n_2)^\beta} \nonumber\\
&&\quad=\;
\sum_{n_1>n_2>0}\frac1{n_1^\alpha}\frac1{(n_1+n_2)^\beta}
+\Big\{
\sum_{n_1,n_2\in\setN}-\sum_{n_2>n_1>0}-\sum_{n_1>0}\delta_{n_2-n_1}
\Big\}\frac1{n_2^\alpha}\frac1{(n_1+n_2)^\beta} 
\nonumber\\&&\quad=\;
\sum_{n_1>n_2>0}\Big[
\frac1{n_1^\alpha}\frac1{(n_1+n_2)^\beta}
+\frac1{n_2^\alpha}\frac1{n_1^\beta}
-\frac1{n_1^\alpha}\frac1{(n_1+n_2)^\beta}
\Big]
-\sum_{n>0} \frac1{n^\alpha}\frac1{(2n)^\beta}
\nonumber\\&&\quad=\;
\zeta(\beta,\alpha)-\frac1{2^\beta}\,\zeta(\alpha+\beta) \;.
\ea
To get to the third line, we have shifted $n_1\rightarrow n_1-n_2$ and $n_1\leftrightarrow n_2$ in the second and third terms 
of the second line, respectively, whereupon a sum that we cannot solve cancels and leaves us with (multiple) zeta values, 
see \eq\nr{eq:doublesum1}. The double sum in the first line of \eq\nr{eq:HH} has the form
\ba
\sum_{n_1>n_2>0}\Big(\frac1{n_1^\alpha}\frac1{n_2^\beta}+\frac1{n_2^\alpha}\frac1{n_1^\beta}\Big)
&=& \zeta(\alpha)\,\zeta(\beta)-\zeta(\alpha+\beta) \;,
\ea
where the solution follows from the shuffle identity \eq\nr{eq:shuffle}.

%
\subsection{Proof of statement (b)}

In a similar way, plugging \eq\nr{eq:newconj} into \eq\nr{eq:Sm}, re-expressing the `unwanted mass' in the prefactor 
in terms of the two others, shifting summation indices appropriately and using the coefficients' symmetry,
\ba
\frac{\Sm_{\eta_3\eta_2\eta_1}^{\nu_3\nu_2\nu_1}}{\intBasic}
&=&
\sum_{j=1-\nu_1}^{\nu_2-1} (-1)^j\,c^{(\Nu)}_{\nu_1,\nu_2;j}(d) 
\sum_{k=0}^{\eta_3} \bin{\eta_3}{k} (-1)^{k-\eta_3}
\sum_{n_1>n_2>0}
n_2^{d-2\Nu+\Eta+\ell_1} n_1^{d-\ell_1}
\nonumber\\&+&
\sum_{j=1-\nu_1}^{\nu_3-1} (-1)^j\,c^{(\Nu)}_{\nu_1,\nu_3;j}(d) 
\sum_{k=0}^{\eta_2} \bin{\eta_2}{k} (-1)^{k-\eta_2}
\sum_{n_1>n_2>0}
(n_1-n_2)^{d-2\Nu+\Eta+\ell_1} n_1^{d-\ell_1}
\nonumber\\&+&
\sum_{j=1-\nu_2}^{\nu_3-1} (-1)^\Nu\,c^{(\Nu)}_{\nu_2,\nu_3;j}(d) 
\sum_{k=0}^{\eta_1} \bin{\eta_1}{k} 
\sum_{n_1>n_2>0}
(n_1-n_2)^{d-2\Nu+\Eta+\ell_2} n_2^{d-\ell_2}
\;,\\
\frac{\Sm_{\eta_3\eta_1\eta_2}^{\nu_3\nu_1\nu_2}}{\intBasic}
&=&
\sum_{j=1-\nu_1}^{\nu_2-1} (-1)^j\,c^{(\Nu)}_{\nu_1,\nu_2;j}(d) 
\sum_{k=0}^{\eta_3} \bin{\eta_3}{k} (-1)^{k}
\sum_{n_1>n_2>0}
n_2^{d-\ell_1} n_1^{d-2\Nu+\Eta+\ell_1}
\nonumber\\&+&
\sum_{j=1-\nu_1}^{\nu_3-1} (-1)^\Nu\,c^{(\Nu)}_{\nu_1,\nu_3;j}(d) 
\sum_{k=0}^{\eta_2} \bin{\eta_2}{k} 
\sum_{n_1>n_2>0}
(n_1-n_2)^{d-2\Nu+\Eta+\ell_1} n_2^{d-\ell_1}
\\&+&
\sum_{j=1-\nu_2}^{\nu_3-1} (-1)^j\,c^{(\Nu)}_{\nu_2,\nu_3;j}(d) 
\sum_{k=0}^{\eta_1} \bin{\eta_1}{k} (-1)^{k-\eta_1}
\sum_{n_1>n_2>0}
(n_1-n_2)^{d-2\Nu+\Eta+\ell_2} n_1^{d-\ell_2} \nonumber
\;.
\ea
The double sums needed here are given in \eqs\nr{eq:doublesum1}, \nr{eq:doublesum2} and \nr{eq:doublesum3}, respectively, 
resulting in single and double zeta values. Combining with the prefactors corresponding to \eq\nr{eq:pos}, we therefore have
\ba
&&
\frac{\Sm_{\eta_3\eta_2\eta_1}^{\nu_3\nu_2\nu_1}+(-1)^{\eta_3}\Sm_{\eta_3\eta_1\eta_2}^{\nu_3\nu_1\nu_2}
+(-1)^{\eta_2}[\Sp_{\eta_2\eta_1\eta_3}^{\nu_2\nu_1\nu_3}+\Sp_{\eta_1\eta_2\eta_3}^{\nu_1\nu_2\nu_3}]}{(-1)^\Nu\,\intBasic} \\
&&=\!\!
\sum_{j=1-\nu_1}^{\nu_2-1} \!\!\! c^{(\Nu)}_{\nu_1,\nu_2;j}(d) 
\sum_{k=0}^{\eta_3} \bin{\eta_3}{k} 
\Big\{
(-1)^{\ell_3}\Big[\zeta(d_1,e_1)\!+\!\zeta(e_1,d_1)\Big]
+(-1)^{\eta_2}\Big[\zeta(e_1)\zeta(d_1)\!-\!\zeta(e_1\!+\!d_1)\Big]
\Big\}
\nonumber\\&&+\!\!
\sum_{j=1-\nu_1}^{\nu_3-1} \!\!\! c^{(\Nu)}_{\nu_1,\nu_3;j}(d) 
\sum_{k=0}^{\eta_2} \bin{\eta_2}{k} 
\Big\{
(-1)^{\ell_2}\Big[\zeta(d_1,e_1)\!+\!\zeta(e_1,d_1)\!-\!\frac{\zeta(e_1\!+\!d_1)}{2^{e_1}}\Big]
+(-1)^{\eta_3} \zeta(e_1)\zeta(d_1)
\Big\}
\nonumber\\&&+\!\!
\sum_{j=1-\nu_2}^{\nu_3-1} \!\!\! c^{(\Nu)}_{\nu_2,\nu_3;j}(d) 
\sum_{k=0}^{\eta_1} \bin{\eta_1}{k} 
\Big\{
\zeta(d_2)\zeta(e_2)
+(-1)^{\ell_2}\Big[(-1)^\Eta\zeta(d_2,e_2)\!+\!\zeta(e_2,d_2)\!-\!\frac{\zeta(e_2\!+\!d_2) }{2^{e_2}}\Big]
\Big\}.\nonumber
\ea
Here, $\ell_p=\Nu-\eta_p-j-k$, $d_p=\ell_p-d$ and $e_p=2\Nu-d-\Eta-\ell_p$.
Note that the prefactor of \eq\nr{eq:pos} guarantees that $\Eta$ is even, 
so we can drop the $(-1)^\Eta$ in the last line above. 
Now, all multiple zeta values occur in symmetric pairs, and can hence be converted to single zeta values 
(and products thereof) by the shuffle relation \eq\nr{eq:shuffle}:
\ba \la{eq:proofb}
&&
\frac{\Sm_{\eta_3\eta_2\eta_1}^{\nu_3\nu_2\nu_1}+(-1)^{\eta_3}\Sm_{\eta_3\eta_1\eta_2}^{\nu_3\nu_1\nu_2}
+(-1)^{\eta_2}[\Sp_{\eta_2\eta_1\eta_3}^{\nu_2\nu_1\nu_3}+\Sp_{\eta_1\eta_2\eta_3}^{\nu_1\nu_2\nu_3}]}{(-1)^\Nu\,\intBasic} \\
&&=\!\!
\sum_{j=1-\nu_1}^{\nu_2-1} \!\!\! c^{(\Nu)}_{\nu_1,\nu_2;j}(d) 
\sum_{k=0}^{\eta_3} \bin{\eta_3}{k} 
\Big\{
\Big[(-1)^{\ell_3}+(-1)^{\eta_2}\Big]
\Big[\zeta(e_1)\zeta(d_1)-\zeta(e_1+d_1)\Big]
\Big\}
\nonumber\\&&+\!\!
\sum_{j=1-\nu_1}^{\nu_3-1} \!\!\! c^{(\Nu)}_{\nu_1,\nu_3;j}(d) 
\sum_{k=0}^{\eta_2} \!\bin{\eta_2}{k}\! 
\Big\{
\!(-1)^{\ell_2}\Big[\zeta(d_1)\zeta(e_1)\!-\!\zeta(e_1\!+\!d_1)\!-\!\frac{\zeta(e_1\!+\!d_1)}{2^{e_1}}\Big]
\!+\!(-1)^{\eta_3} \zeta(e_1)\zeta(d_1)
\Big\}
\nonumber\\&&+\!\!
\sum_{j=1-\nu_2}^{\nu_3-1} \!\!\! c^{(\Nu)}_{\nu_2,\nu_3;j}(d) 
\sum_{k=0}^{\eta_1} \bin{\eta_1}{k} 
\Big\{
\zeta(d_2)\zeta(e_2)
+(-1)^{\ell_2}\Big[\zeta(d_2)\zeta(e_2)-\zeta(e_2+d_2)-\frac{\zeta(e_2+d_2)}{2^{e_2}}\Big]
\Big\}.\nonumber
\ea

%
\subsection{Proof of statement (c)}

Looking at the structure of \eq\nr{eq:proofb}, we are happy with the products of zetas (which we can in the end interpret as 
products of 1-loop sum-integrals \eq\nr{eq:1loop}), and note that all remaining single zeta values have arguments 
$e_p+d_p=2\Nu-2d-\Eta$ that do not depend on the summation indices $j$, $k$. 
For these terms, the $k$-sum can be easily performed using
\ba
\sum_{k=0}^N \bin{N}{k} x^k &=& (1+x)^N \;,\quad
\sum_{k=0}^N \bin{N}{k} (-1)^k = \delta_N \;,
\ea
producing
\ba 
&&
-\frac{\Sm_{\eta_3\eta_2\eta_1}^{\nu_3\nu_2\nu_1}+(-1)^{\eta_3}\Sm_{\eta_3\eta_1\eta_2}^{\nu_3\nu_1\nu_2}
+(-1)^{\eta_2}[\Sp_{\eta_2\eta_1\eta_3}^{\nu_2\nu_1\nu_3}+\Sp_{\eta_1\eta_2\eta_3}^{\nu_1\nu_2\nu_3}]}
{(-1)^\Nu\,\intBasic\,\zeta(2\Nu-2d-\Eta)} \Big|_{\rm{single~zetas}}\\
&&\qquad=
\sum_{j=1-\nu_1}^{\nu_2-1} c^{(\Nu)}_{\nu_1,\nu_2;j}(d) 
\Big[(-1)^{\Nu-\eta_3-j}\delta_{\eta_3}+(-1)^{\eta_2}2^{\eta_3}\Big]
\nonumber\\&&\qquad+
\sum_{j=1-\nu_1}^{\nu_3-1} c^{(\Nu)}_{\nu_1,\nu_3;j}(d) 
\Big[(-1)^{\Nu-\eta_2-j}\delta_{\eta_2}+(-1)^{\Nu-\eta_2-j}(\tfrac12)^{\Nu-d-\eta_3+j}\Big]
\nonumber\\&&\qquad+
\sum_{j=1-\nu_2}^{\nu_3-1} c^{(\Nu)}_{\nu_2,\nu_3;j}(d) 
\Big[(-1)^{\Nu-\eta_2-j}\delta_{\eta_1}+(-1)^{\Nu-\eta_2-j}(\tfrac12)^{\Nu-d-\eta_3+j}\Big]
\;.\nonumber
\ea
Comparing with \eqs\nr{eq:new0mm}--\nr{eq:newr32}, the remaining sums over $j$ can now be seen to be nothing but 
the coefficients of special cases of $\intB$, 
\ba \la{eq:proofc}
&&
\Sm_{\eta_3\eta_2\eta_1}^{\nu_3\nu_2\nu_1}+(-1)^{\eta_3}\Sm_{\eta_3\eta_1\eta_2}^{\nu_3\nu_1\nu_2}
+(-1)^{\eta_2}[\Sp_{\eta_2\eta_1\eta_3}^{\nu_2\nu_1\nu_3}+\Sp_{\eta_1\eta_2\eta_3}^{\nu_1\nu_2\nu_3}] \Big|_{\rm{single~zetas}}\\
&&=
-(-1)^\Nu\,\intBasic\,\zeta(2\Nu-2d-\Eta) \Big\{
(-1)^{\Nu-\eta_3}\delta_{\eta_3} \beta^{\nu_3\nu_1\nu_2}
+(-1)^{\eta_2}2^{\eta_3} (-1)^\Nu \hr_1^{\nu_1\nu_2\nu_3}
\nonumber\\&&\quad+
(-1)^{\Nu-\eta_2}\delta_{\eta_2} \beta^{\nu_2\nu_1\nu_3}
+(-1)^{\Nu-\eta_2}\delta_{\eta_1} \beta^{\nu_1\nu_2\nu_3}
+(-1)^{\Nu-\eta_2}(\tfrac12)^{\Nu-d-\eta_3} \frac{2^\Nu}{4}\hr_2^{\nu_1\nu_2\nu_3}
\Big\}
\nonumber\\&&=
-\zeta(2\Nu-2d-\Eta) \Big[
\delta_{\eta_3} \intB_{0,1,1}^{\nu_3\nu_1\nu_2}
+\delta_{\eta_2} \intB_{0,1,1}^{\nu_2\nu_1\nu_3}
+(-1)^{\eta_2}\delta_{\eta_1} \intB_{0,1,1}^{\nu_1\nu_2\nu_3}
+(-1)^{\eta_2}2^{\eta_3}\intB_{1,1,2}^{\nu_1\nu_2\nu_3}
\Big]
\;,\nonumber
\ea
which (after rewriting $(-1)^{\eta_2}\delta_{\eta_1}=(-1)^\Eta(-1)^{\eta_3}\delta_{\eta_1}$ and noticing that $\Eta$ is 
even due to the prefactor of \eq\nr{eq:pos}) cancels exactly against the second line of \eq\nr{eq:pos}. 
This completes the proof that only products of single zeta values remain in the full result for our sum-integral $\sumintL$.

%
\subsection{Proof of statement (d)}

All that we are left with for the curly brackets of \eq\nr{eq:pos} are the products of zeta values in \eq\nr{eq:proofb}.
Using that $\Eta$ is even (or $(-1)^\Eta=1$), we can re-write the specific combinations of prefactors in a 
more convenient form, to finally obtain ($\ell_p\equiv \Nu-\eta_p-j-k$)
\ba \la{eq:fin}
\sumintL_\nabc^\eabc(d,T) &=&
\frac{T^2[1+(-1)^\Eta]}{(2\pi T)^{2\Nu-\Eta-2d}}\,\intBasic \,(-1)^\Nu \,\times
\\&\Big\{&
\sum_{j=1-\nu_1}^{\nu_2-1} \!\!\! c^{(\Nu)}_{\nu_1,\nu_2;j}(d) 
\sum_{k=0}^{\eta_3} \!\bin{\eta_3}{k} (-1)^{\eta_2}\, [1+(-1)^{\ell_1}]\,\zeta(\ell_1\!-\!d)\,\zeta(2\Nu\!-\!\Eta\!-\!\ell_1\!-\!d)
\nonumber\\&+&
\sum_{j=1-\nu_1}^{\nu_3-1} \!\!\! c^{(\Nu)}_{\nu_1,\nu_3;j}(d) 
\sum_{k=0}^{\eta_2} \!\bin{\eta_2}{k} (-1)^{\eta_3}\, [1+(-1)^{\ell_1}]\,\zeta(\ell_1\!-\!d)\,\zeta(2\Nu\!-\!\Eta\!-\!\ell_1\!-\!d)
\nonumber\\&+&
\sum_{j=1-\nu_2}^{\nu_3-1} \!\!\! c^{(\Nu)}_{\nu_2,\nu_3;j}(d) 
\sum_{k=0}^{\eta_1} \!\bin{\eta_1}{k}  [1+(-1)^{\ell_2}]\,\zeta(\ell_2-d)\,\zeta(2\Nu-\Eta-\ell_2-d)
\Big\} \;. \nonumber
\ea
We recall from \eq\nr{eq:BnormJ} that the normalization factor $\intBasic=\frac{\Gamma^2(1-d/2)}{(4\pi)^{d}}=[J_1(d,1)]^2$ 
is a 1-loop massive tadpole squared (taken at $m=1$), cf.\ \eq\nr{eq:1loopJ}.
Note that the factors $[1+(-1)^\Eta]$ and $[1+(-1)^{\ell_p}]$ ensure that only $\zeta(n_{\rm even}-d)$ 
at even integers $n$ 
contribute to the right-hand side of \eq\nr{eq:fin}.

While the symmetries of the integral $\sumintL$ are not made explicit in \eq\nr{eq:fin}, it is easy to verify that they are indeed satisfied.
For example, interchanging the first and second set of indices, $(\nu_1,\eta_1)\leftrightarrow(\nu_2,\eta_2)$,
induces an overall sign depending on the parity of $\eta_3$, see \eq\nr{eq:symms}.  
To check this, one first observes that under this index permutation the roles of two last two lines interchange, 
while for the remaining line in the curly brackets one applies the first of \eq\nr{eq:csy} 
followed by a change of summation indices $j\rightarrow-j$ and $k\rightarrow\eta_3-k$.

%
\section{Result and extension}
\la{se:result}

Let us finally re-write our result as products of 1-loop sum-integrals $I_\nu^\eta(d,T)$ from \eq\nr{eq:1loop}. 
Each of the zeta functions in \eq\nr{eq:fin} will map onto one such 1-loop sum-integral $I_\nu^\eta(d,T)$, 
and we need to distinguish the cases $\zeta(2n-d)$ with $n>0$ 
(giving master sum-integrals $I_{n}^0$ without numerator) 
and $n\le0$ (giving master sum-integrals $I_1^{2-2n}$ with numerators). 
Both cases can be covered by the single expression
\ba \la{eq:ihat}
\zeta(2n\!-\!d) =
\frac{(2\pi T)^{2n-d}}{T\,J_1(d,1)}\,
\hat{I}_n(d,T)
\,,\;\;
\hat{I}_n(d,T) \equiv
\frac{\Gamma(\sigma_{n})\,I_{\sigma_{n}}^{2\sigma_{n}-2n}(d,T)}{2\,\po{1-\tfrac{d}2}{\sigma_{n}-1}}
\,,\;\;
\sigma_{n} \equiv {\rm max}(n,1)\,,\quad
\ea
with $J_1(d,1)$ the 1-loop tadpole of \eq\nr{eq:1loopJ}.
Doing these identifications, we get (recall the abbreviations $\Nu=\nu_1+\nu_2+\nu_3$, $\Eta=\eta_1+\eta_2+\eta_3$ 
and $\ell_p=\Nu-\eta_p-j-k$)
\ba\la{eq:final}
\sumintL_\nabc^\eabc(d,T) &=&
[1+(-1)^\Eta] \,(-1)^\Nu \,\times
\nonumber\\&\Big\{&
\!\!\!\sum_{j=1-\nu_1}^{\nu_2-1} \!\!\! c^{(\Nu)}_{\nu_1,\nu_2;j}(d) 
\sum_{k=0}^{\eta_3} \!\bin{\eta_3}{k} (-1)^{\eta_2}\, [1\!+\!(-1)^{\ell_1}]\,
\hat{I}_{{\ell_1}/2}(d,T)\,\hat{I}_{({2\Nu-\Eta-\ell_1})/2}(d,T)
\nonumber\\&+&
\!\!\!\sum_{j=1-\nu_1}^{\nu_3-1} \!\!\! c^{(\Nu)}_{\nu_1,\nu_3;j}(d) 
\sum_{k=0}^{\eta_2} \!\bin{\eta_2}{k} (-1)^{\eta_3}\, [1\!+\!(-1)^{\ell_1}]\,
\hat{I}_{{\ell_1}/2}(d,T)\,\hat{I}_{({2\Nu-\Eta-\ell_1})/2}(d,T)
\nonumber\\&+&
\!\!\!\sum_{j=1-\nu_2}^{\nu_3-1} \!\!\! c^{(\Nu)}_{\nu_2,\nu_3;j}(d) 
\sum_{k=0}^{\eta_1} \!\bin{\eta_1}{k}  [1\!+\!(-1)^{\ell_2}]\,
\hat{I}_{{\ell_2}/2}(d,T)\,\hat{I}_{({2\Nu-\Eta-\ell_2})/2}(d,T)
\Big\} ,\quad
\ea
which, together with \eq\nr{eq:kallen3}, constitutes our final result. Sample Mathematica code that implements 
our final result is given in \app\ref{app:code}.

%
\subsection{Non-positive indices and full reduction algorithm}

Having the decomposition \eq\nr{eq:final} for scalar 
bosonic two-loop vacuum sum-integrals $\sumintL_\nabc^\eabc(d,T)$ with positive propagator powers $\nu_i\in\setN$ at hand, 
we can easily extend it to be valid for all non-negative integers $\nu_i\in\setN_0$ by analyzing the boundary cases
when (at least) one $\nu_i=0$. Analyzing \eq(3.24) of \cite{Davydychev:2022dcw} with boundary conditions \eqs(2.7)--(2.9)
therein, it is sufficient to amend the definition of the coefficient functions given in \eq\nr{eq:kallen3} above with the specific cases
\ba \la{eq:bc}
c^{(\nu_a+\nu_b)}_{\nu_a,\nu_b;\nu_b-\nu_a}(d) &\equiv& (-1)^{\nu_a+\nu_b}
\frac{\po{1-\frac{d}2}{\nu_a-1}}{\Gamma(\nu_a)}
\frac{\po{1-\frac{d}2}{\nu_b-1}}{\Gamma(\nu_b)} \;,
\ea
while the remaining boundary coefficients ($c^{(\Nu)}_{\nu_a,0;j}=0=c^{(\Nu)}_{0,\nu_b;j}$ 
and $c^{(\nu_a+\nu_b)}_{\nu_a,\nu_b;j\neq \nu_b-\nu_a}=0$)
are already correctly reproduced by \eq\nr{eq:kallen3}.

With a little bit of additional work, a further useful extension to {\em all} integer powers $\nu_i\in\setZ$ can be done as follows.
Given an $\sumintL_\nabc^\eabc(d,T)$ as defined in \eq\nr{eq:L}, with $\eta_i\in\setN_0$ and $\nu_i\in\setZ$
with at least one non-positive index $\nu_i\le0$, apply 
\ba \la{eq:symms}
L_\nabc^\eabc(d,T)\;=\;L_{\nu_1,\nu_3,\nu_2}^{\eta_1,\eta_3,\eta_2}(d,T)
\quad,\quad
L_\nabc^\eabc(d,T)\;=\;(-1)^{\eta_3}\,L_{\nu_2,\nu_1,\nu_3}^{\eta_2,\eta_1,\eta_3}(d,T)
\ea
(following from momentum shifts $Q\rightarrow P-Q$ and $P\leftrightarrow Q$, respectively)
to order the lower indices such that $\nu_1\ge\nu_2\ge\nu_3$.
Then, simplify and factorize those cases by 
\\(a) realizing that integrals with odd upper-index sum $\Eta=$odd vanish due to the denominator's 
invariance under the shift $\{P,Q\}\rightarrow\{-P,-Q\}$;
\\(b) if $\eta_3>0$, expanding the corresponding numerator binomially as in \eq\nr{eq:check2};
\\(c) if $\nu_3<0$, expand the corresponding numerator double-binomially to map onto 1-loop tensor tadpoles 
$T_{s,t}^{\mu_1\dots\mu_n}=\Tinti{P}\frac{(U\cdot P)^t}{[P^2]^s}\,P^{\mu_1}\cdots P^{\mu_n}$, 
use the 1-loop tensor tadpole decomposition into orthogonal bases made out of $U^\mu=(1,\bf{0})$ and $g^{\mu\nu}$ 
as given in \cite{Navarrete:2022adz} and compute the basis tensor contractions;
leading to 
\ba \la{eq:negNu3}
\sumintL_{\nabc\le0}^\eabc(d,T) &=& 
\frac12[1+(-1)^\Eta]
\sum_{j=0}^{\eta_3}
\sum_{n=0}^{-\nu_3}\sum_{k=0}^{-\nu_3-n} 
\sum_{\ell=0}^{\floor{n/2}} \frac{(-1)^j\,\bin{\eta_3}{j}\,(-2)^n\,(-\nu_3)!}{4^\ell\,\ell!\,(n-2\ell)!\,k!\,(-\nu_3-n-k)!}
\;\times\nonumber\\&\times&
\frac{\po{\frac{d}2}{\ell}\; I_{\nu_1+\nu_3+n+k}^{\eta_1+\eta_3-j+n}(d,T)\cdot I_{\nu_2-k}^{\eta_2+j+n}(d,T)}
{\po{\frac{d}2+1-\nu_2+k}{\ell}\po{\frac{d}2+1-\nu_1-\nu_3-k-n}{\ell}}\;.
\ea

The 1-loop tadpoles $I$ in \eq\nr{eq:negNu3} can finally be mapped onto the basis $\hat I$ via \eqs\nr{eq:1loop} and \nr{eq:ihat}.
On the one hand, this introduces Gamma factors in the denominators which correctly set all $I_{\nu\le0}^\eta=0$.
To exclude such vanishing terms from the outset, we can also modify the limits of the $k$-sum as
$\sum_{k={\rm max}(0,1-\nu_1-\nu_3-n)}^{{\rm min}(-\nu_3-n,\nu_2-1)}$.
On the other hand, the map introduces prefactors which implement that $I_\nu^{\eta={\rm odd}}=0$.
Modifying the $j$-sum as $\sum_{j=0}^{\eta_3}f(j)\rightarrow\sum_{j=x}^{\floor{(\eta_3+x)/2}}f(2j-x)$ where $x={\rm mod}(\eta_2+n,2)$
then eliminates the need for such explicit prefactors $[1+(-1)^\eta]$ stemming from \eq\nr{eq:1loop}. 
After some slight simplifications, this leaves us with a representation on par with \eq\nr{eq:final}:
\ba \la{eq:finalZ}
\sumintL_{\nabc\le0}^\eabc(d,T) &=& 
\frac12[1+(-1)^\Eta]
\sum_{n=0}^{-\nu_3}
\sum_{j=x}^{\floor{(\eta_3+x)/2}}
\sum_{k={\rm max}(1,n+\Nu-\nu_1)}^{{\rm min}(\nu_2,n+\Nu-1)}
\sum_{\ell=0}^{\floor{n/2}} 
\;\times\nonumber\\&\times&
\frac{(-1)^{\eta_2}\,(-\nu_3)!\,2^{2+n-2\ell}\,\bin{\eta_3}{2j-x}}
{\ell!\,(n-2\ell)!\,(\nu_2-k)!\,(\nu_1-z)!\,\Gamma(z)\,\Gamma(k)}
\;\times\nonumber\\&\times&
\frac{\po{\frac{d}2}{\ell}\, (1-\frac{d}2)_{z-1}\, (1-\frac{d}2)_{k-1}}
{ \po{\frac{d}2+1-z}{\ell} \, \po{\frac{d}2+1-k}{\ell} }\;
\hat{I}_{\Nu-\Eta/2-y}(d,T) \cdot \hat{I}_y(d,T) \;,
\ea
where $x\equiv{\rm mod}(\eta_2+n,2)$
as well as $y\equiv k-j-\floor{(\eta_2+n)/2}$
and $z\equiv\Nu+n-k$.
As a check, we note that for $\nu_3=0$ and positive values of $\nu_1,\nu_2$, \eq\nr{eq:finalZ} as well as \eq\nr{eq:final} with \eq\nr{eq:bc} 
both reduce to single sums\footnote{In particular, the quadruple sum of \eq\nr{eq:finalZ} 
collapses to a single sum since only the terms at $n=0$, $k=\nu_2$ and $\ell=0$ contribute; 
while for \eq\nr{eq:final}, only the first of the three double-sums contributes, with its $j$-sum collapsing to a single term $j=\nu_2-\nu_1$.}, 
and agree in their factorization of $L_{\nu_1,\nu_2,0}^\eabc(d,T)$. 
We can therefore use the much more general \eq\nr{eq:finalZ} for reducing sum-integrals involving non-positive indices, 
without ever applying \eq\nr{eq:bc}.
In order to check \eq\nr{eq:finalZ}, we have verified it against a pre-existing in-house IBP code,
for a large number of integer values for $\nu_i$ and $\eta_i$, finding complete agreement.

We are now in a position to formulate an algorithm that maps every massless scalar 
bosonic two-loop vacuum sum-integral to its 1-loop factorized form:
\\[2mm]Given an $\sumintL_\nabc^\eabc(d,T)$ as defined in \eq\nr{eq:L}, with $\eta_i\in\setN_0$ and $\nu_i\in\setZ$, 
\begin{enumerate}
\item If all $\nu_i$ are positive integers, apply \eq\nr{eq:final} to map $\sumintL\rightarrow\sum \hat{I}\cdot \hat{I}$,
where the $\hat{I}$ are elements of the 1-loop basis defined in \eq\nr{eq:ihat}.
\item If any $\nu_i\le0$, put the lower indices into descending order by \eq\nr{eq:symms} 
and then apply \eq\nr{eq:finalZ} to map $\sumintL\rightarrow\sum \hat{I}\cdot \hat{I}$.
\end{enumerate}
For sample Mathematica code implementing this strategy, see \app\ref{app:code}.

%
\subsection{Checks and examples}
\la{se:checks}

For checks, let us reproduce some known IBP results, using our main result \eq\nr{eq:final}. 
There are short IBP proofs in \eq(5) of \cite{Schroder:2008ex} (see also \eq(A.3) of \cite{Nishimura:2012ee}) 
and \eq(E.2) of \cite{Ghisoiu:2012kn}, respectively, of the (propagator-) weight 3 examples
\ba\la{eq:zero}
\sumintL_{1,1,1}^{0,0,0} &=& 0 \;,\quad \sumintL_{2,1,1}^{2,0,0} \;=\; 0 \;.
\ea
In \app B of \cite{Ghisoiu:2012yk}, we gave the IBP reductions (at weight 3, 4, 5 and 7, respectively) 
\ba
\sumintL_{2,1,1}^{0,2,0} &=& \frac{(d-3)}{(d-5)}\,I_2^0 \, I_1^0 \;,\quad
\sumintL_{3,1,1}^{2,2,0} \;=\; -\frac{(d-4)(d^2-8d+19)}{4(d-7)(d-5)}\, I_2^0 \, I_1^0 \;,\la{eq:L31122}\\
\sumintL_{2,1,1}^{0,0,0} &=& -\frac{I_2^0 \, I_2^0}{(d-5)(d-2)} \;,\\
\sumintL_{3,1,1}^{0,0,0} &=& -\frac{4\, I_3^0 \, I_2^0}{(d-7)(d-2)}  \;,\;\;
\sumintL_{2,2,1}^{0,0,0} \;=\; 0 \;,\;\;
\sumintL_{3,1,2}^{1,1,0} \;=\; \frac{(d-6)(d-5)(d-3)}{2(d-9)(d-7)(d-2)}\, I_3^0 \, I_2^0 \;,\qquad\\
\sumintL_{3,3,1}^{0,0,0} &=& -\frac{12(d-8)(d-5)\, I_4^0 \, I_3^0}{(d-11)(d-9)(d-4)(d-2)} \;.
\ea
Using the Mathematica code of \app\ref{app:code}, we confirm all IBP reductions listed above.

As another check, we can expand the numerator of \eq\nr{eq:L} for positive values of the index $\eta_3$, and 
verify that each $\sumintL$ can be expressed in terms of instances with $\eta_3=0$ as
\ba\la{eq:check2}
\sumintL_\nabc^\eabc(d,T) &=& \sum_{j=0}^{\eta_3} \bin{\eta_3}{j}\,(-1)^j\,\sumintL_\nabc^{\eta_1+\eta_3-j,\eta_2+j,0}(d,T) \;.
\ea
Indeed, testing \eq\nr{eq:check2} for a wide range of integers $\{\nabc,\eabc\}$ by applying \eq\nr{eq:final} on both sides, 
we find perfect agreement. 

Going beyond published results, one could now use \eq\nr{eq:final} to generate a large table of sum-integral reductions, 
to facilitate future calculations and/or serve as a benchmark. We refrain from doing so, however, but list a couple of
interesting cases that exhibit reductions to linear combinations of different master integrals, a feature not yet seen in the
examples above. A few of these multi-term cases (at weight 4, 6 and 8) are 
\ba
\sumintL_{1,1,4}^{2,2,0} &=&
-\frac{(d-4)\,I_2^0\,I_2^0}{4(d-9)(d-7)(d-2)}
-\frac{(d-5)(d-4)\,I_1^0\,I_3^0}{2(d-9)(d-7)}
+\frac{(2d-13)\,I_1^2\,I_4^0}{(d-9)}
\;,\nonumber\\
\sumintL_{1,1,4}^{0,0,0} &=&
-\frac{4 \,I_3^0\,I_3^0}{(d-9) (d-7) (d-4) (d-2)}
-\frac{6  \,I_2^0\,I_4^0}{(d-9) (d-2)} \;,
\nonumber\\\sumintL_{2,2,4}^{0,0,0} &=&
\frac{36 \left(7 d^2-105 d+382\right)  \,I_4^0\,I_4^0}{(d-13) (d-11) (d-9) (d-6) (d-4) (d-2)}
+\frac{16 (d-10) (d-5)  \,I_3^0\,I_5^0}{(d-13) (d-11) (d-4) (d-2)} \;.
\nonumber
\ea
On the other hand, from \eqs\nr{eq:symms} and \nr{eq:negNu3} some examples for non-positive indices read
\ba
\sumintL_{1,2,-2}^{2,1,1} &=& 
2(d-3)\,I_1^2\,I_1^4 
\;,\nonumber\\
\sumintL_{2,-3,3}^{1,2,3} &=& 
(d^3-14d^2+62d-94)\,I_1^2\,I_1^4 
+3(d-2)\,I_1^0\,I_1^6 
\;,\nonumber\\
\sumintL_{-4,2,3}^{2,5,3} &=& 
8(d^2-4d+10)\,I_1^6\,I_1^6 
-2(d-2)(d^2-6d+33)\,I_1^4\,I_1^8
+4(d-2)(d-1)\,I_1^2\,I_1^{10} 
\;.\nonumber
\ea

%
\section{Conclusions and outlook}
\la{se:conclu}

We have presented an analytic solution for two-loop scalar massless bosonic vacuum sum-integrals $\sumintL$,
with general integer values of propagator and numerator powers. 
Our key result presented in \eq\nr{eq:final} provides a mapping to analytically known one-loop sum-integrals,
based on a similar decomposition that had been derived earlier for a class of closely related massive two-loop
continuum integrals \cite{Davydychev:2022dcw}.

While some fixed-index cases of such a two-loop onto one-loop sum-integral mapping had already been observed 
in the literature, our general-index proof exposes the mechanism behind this most helpful mathematical fact, 
and allows to eliminate all such sum-integrals from perturbative expansions, on an algebraic level. 

For convenience, we have extended our result to be applicable also for non-positive propagator indices,
essentially eliminating the need for tensor reduction when inverse propagators are present in the numerator.
In practice, the reader might find the algorithmic implementation presented in \app\ref{app:code} useful,
where all cases are combined.

Going beyond the scalar sum-integrals discussed here, one natural future generalization would be 
to consider tensors $P^{\mu_1}\cdots P^{\mu_k}\,Q^{\mu_{k+1}}\cdots Q^{\mu_n}$ in the numerator
of \eq\nr{eq:L}. If such tensors are fully symmetric in the $\mu_i$, one can trivially apply
the decomposition presented in \eqs(8), (16) of \cite{Navarrete:2022adz}. For the general case, however,
some additional work is required. This particular generalization might become interesting when 
evaluating four-loop vacuum sum-integrals, since then product-like structures 
$\sim\sumintL\cdot\sumintL$ appear, with a numerator potentially 
containing scalar products between momenta from different $\sumintL$, necessitating a tensor decomposition. 
In practice, for fixed-index cases this can also be achieved by IBP relations, such that we postpone the 
analytic generalization to future work.

Other interesting avenues for future work would be generalizations to massless fermions (with and without chemical potential $\muf$), 
and higher loops. 
Including (bosonic as well as fermionic) propagator masses deserves to be investigated as well, but seems to be a different game altogether. 
Note that not even the massive one-loop sum-integral is known analytically; 
what is known are only high-temperature expansions (as presented in textbooks, such as in the appendix of \cite{Kapusta:2023eix}) 
or practical integral representations and evaluations of some terms in their $\ep$-expansion (see e.g.\ \app A of \cite{Laine:2019uua} 
as well as \cite{Laine:2017hdk}).  

As for sum-integral factorization properties at higher loops, there seems to be little hope,
if our observations of some particular fixed-index cases of higher-loop IBP reductions are of any guidance.
One could hope, however, that for some special integral sectors (i.e.\ particular propagator subsets)
such general simplifications might exist.

%
\acknowledgments

We wish to thank I.~Kondrashuk for discussions at early stages of this work,
and L.~Gil for checking our Mathematica code.
The work of A.D.\ was partially supported by CONICYT PCI/MEC 80180071.
P.N.\ has been supported by an ANID grant Mag\'ister Nacional No.\ 22211544, and 
by the Academy of Finland grants No.\ 347499, 353772 and 1322507.
Y.S.\ acknowledges support from ANID under FONDECYT projects No.\ 1191073 and 1231056.

%
\appendix

%
\section{Single and double sums}
\la{se:sums}

Other than single sums
\ba
\sum_{n=1}^\infty \frac1{n^s} &=& \zeta(s) \;,
\ea
we encounter double sums of the following types:
\ba
\la{eq:doublesum1}
\sum_{n_1>n_2>0} \frac1{n_1^{s_1}\,n_2^{s_2}} &\equiv& \zeta(s_1,s_2) \;,\\
\la{eq:doublesum2}
\sum_{n_1>n_2>0} \frac1{n_1^{s_1}\,(n_1-n_2)^{s_2}} &=& \zeta(s_1,s_2) \;,\\
\la{eq:doublesum3}
\sum_{n_1>n_2>0} \frac1{(n_1-n_2)^{s_1}\,n_2^{s_2}} &=& \zeta(s_1)\,\zeta(s_2) \;.
\ea
For the multiple zeta function $\zeta(s_1,s_2)$ occurring above, the shuffle identity (see also \fig\ref{figB})
\ba
\sum_{n_1=1}^\infty\,\sum_{n_2=1}^\infty \frac1{n_1^{s_1}\,n_2^{s_2}} &=& 
\Big(\sum_{n_1>n_2>0}+\sum_{n_2>n_1>0}+\sum_{n_1=1}^\infty\delta_{n_2-n_1}\Big)
\frac1{n_1^{s_1}\,n_2^{s_2}} \\
\la{eq:shuffle}
\Leftrightarrow\qquad\quad
\zeta(s_1)\,\zeta(s_2) &=& \zeta(s_1,s_2)+\zeta(s_2,s_1)+\zeta(s_1+s_2) 
\ea
can be used to sort the two arguments. 

We also encounter two types of unknown double sums
\ba\la{eq:doublesum4}
\sum_{n_1>n_2>0} \frac1{(n_1+n_2)^{s_1}\,n_2^{s_2}} &\equiv& \sumZ(s_1,s_2) \;,\\
\la{eq:doublesum5}
\sum_{n_1>n_2>0} \frac1{(n_1+n_2)^{s_1}\,n_1^{s_2}} &=& \zeta(s_1,s_2)-\sumZ(s_1,s_2)-\frac1{2^{s_1}}\,\zeta(s_1+s_2) \;,
\ea
which in \eq\nr{eq:pos} fortunately always occur in combinations where the unknown part $\sumZ$ drops out, as proven 
in \se\ref{se:proofA}.

\figB

%
\section{Code snippet: Reduction of 2-loop sum-integral $\sumintL$}
\la{app:code}

\newcommand{\ttn}{{\tt n}}
\newcommand{\tte}{{\tt e}}

Here, we show sample Mathematica code that implements our \eqs\nr{eq:ihat}, \nr{eq:final}, \nr{eq:symms} and \nr{eq:finalZ}.
It provides the six-index function ${\tt L}[\ttn_1,\ttn_2,\ttn_3,\tte_1,\tte_2,\tte_3]=\sumintL_{n_1,n_2,n_3}^{e_1,e_2,e_3}(d,T)$ 
which evaluates to a linear combination of products of one-loop master sum-integrals ${\tt i[s,t]}=I_s^t(d,T)$, 
with coefficients that are rational functions in the number of spatial dimensions $d$.
\begin{verbatim}
ClearAll[d,i,c,h,L];
c[nu_,na_,nb_,j_] := Module[ {nj=Ceiling[(nu+j)/2], nc=nu-na-nb}, 
     (-1)^(nu-nj+1) Pochhammer[1-d/2,nj-j-1] / (2 Pochhammer[1/2,nj-nb-nc] 
     Pochhammer[1/2,nj-j-na-nc] (nc-1)!) Sum[ Pochhammer[d/2-nj+1,k-1] 
     (nj-k-1)! / (Pochhammer[d/2+3/2-nu,nj-k] Pochhammer[1/2,nc-nj+k] 
     (k-1)! (k-j-1)! (nb-k)! (na-k+j)!), {k,Max[1+j,1],Min[na+j,nb]}] ];
h[n_] := Module[ {s=Max[n,1]}, 
     Gamma[s] i[s,2s-2n]/2/Pochhammer[1-d/2,s-1] ];
L[n1_,n2_,n3_,e1_,e2_,e3_] := Which[ n1<n2, (-1)^e3 L[n2,n1,n3,e2,e1,e3],
     n2<n3, L[n1,n3,n2,e1,e3,e2], True, Collect[ Module[ {nu=n1+n2+n3, 
     eta=e1+e2+e3}, (1+(-1)^eta) If[ n3>0, 
     (*all indices n1..n3 positive: apply 6.2*) (-1)^nu*(
     Sum[ c[nu,n1,n2,j] Sum[ With[ {el=nu-e1-j-k}, Binomial[e3,k] (-1)^e2 
     (1+(-1)^el) h[el/2] h[nu-eta/2-el/2] ], {k,0,e3}], {j,1-n1,n2-1}] +
     Sum[ c[nu,n1,n3,j] Sum[ With[ {el=nu-e1-j-k}, Binomial[e2,k] (-1)^e3 
     (1+(-1)^el) h[el/2] h[nu-eta/2-el/2] ], {k,0,e2}], {j,1-n1,n3-1}] +
     Sum[ c[nu,n2,n3,j] Sum[ With[ {el=nu-e2-j-k}, Binomial[e1,k] 
     (1+(-1)^el) h[el/2] h[nu-eta/2-el/2] ], {k,0,e1}], {j,1-n2,n3-1}] ), 
     (*not all indices n1..n3 positive: apply 6.6*)
     2 (-1)^e2 (-n3)! Sum[ With[ {x=Mod[e2+n,2]}, 
     Sum[ With[ {y=k-j-Floor[(e2+n)/2], z=nu+n-k}, 
     Binomial[e3,2j-x] 2^n/(l!(n-2l)!(n2-k)!(n1-z)!Gamma[z]Gamma[k]4^l) 
     Pochhammer[d/2,l] Pochhammer[1-d/2,z-1] Pochhammer[1-d/2,k-1]/
     (Pochhammer[d/2+1-z,l] Pochhammer[d/2+1-k,l]) h[nu-eta/2-y] h[y]], 
     {j,x,Floor[(e3+x)/2]}, {k,Max[1,n+nu-n1],Min[n2,n+nu-1]}, 
     {l,0,Floor[n/2]} ] ], {n,0,-n3}]] ], i[__], Factor ] ];
\end{verbatim}
After loading these definitions, one can simply call the function {\tt L} to obtain the factorized version of the corresponding sum-integral.
As an example, the result for $\sumintL_{3,1,1}^{2,2,0}(d,T)$ shown in \eq\nr{eq:L31122} is generated by executing
\begin{verbatim}
(*example use*)
L[3,1,1,2,2,0]
\end{verbatim}
Note added: We thank Luis Gil for pointing that the overall factor $(-1)^\Nu$ from \eq\nr{eq:final} was missing in the Mathematica code. We have corrected this now in the above code.

%

\end{document}